\documentstyle[editedvolume]{crckapb}
\def\ltwid{\mathrel{\raise.3ex\hbox{$<$\kern-.75em\lower1ex\hbox{$\sim$}}}}

\begin{opening}
\author{P.M. LUBIN}
\title{DEGREE SCALE ANISOTROPY:\hfil\break CURRENT STATUS
}
\institute{University of California Santa Barbara\\
		   Physics Department \\  Santa Barbara, CA  93106, USA \\
		e-mail: lubin@cfi.ucsb.edu}
\institute{Center for Particle Astrophysics\\
	   University of California\\  Berkeley, CA 94720, USA}
\end{opening}
\runningtitle{DEGREE SCALE ANISOTROPY}

\begin{document}

\section{ABSTRACT}

The Cosmic Background Radiation gives us one of the few probes into the
density perturbations in the early universe that should later lead to the
formation of structure we now observe. Recent advances in degree scale
anisotropy measurements have allowed us to begin critically testing
cosmological models. Combined with the larger scale measurements from COBE
we are now able to directly compare data and theory. These measurements
promise future progress in understanding structure formation. Because of the
extreme sensitivities needed (1-10 ppm) and the difficulties of foreground
sources, these measurements require not only technological advances in
detector and measurement techniques, but multi spectral measurements and
careful attention to low level systematic errors. This field is advancing
rapidly and is in a true discovery mode. Our own group has been involved in
a series of eleven experiments over the last six years using the ACME
(Advanced Cosmic Microwave Explorer) payload which has made measurements at
angular scales from 0.3 to 3 degrees and over a wavelength range from 1 to
10 mm. The recent data from these and other measurements will be reviewed as
well as some of the challenges and potential involved in these and future
measurements.

\section{Introduction}

The Cosmic Background Radiation (CBR) provides a unique opportunity to test
cosmological theories. It is one of the few fossil remnants of the early
universe to which we have access at the present. Spatial anisotropy
measurements of the CBR in particular can provide a probe of density
fluctuations in the early universe. If the density fluctuation spectrum can
be mapped at high redshift, the results can be combined with other
measurements of large scale structure in the universe to provide a coherent
cosmological model.

\begin{figure}
\vspace{11cm}  
\caption{Theoretical CBR Power Spectrum (model results from
P. Steinhardt and R. Bond, private communication for a $Q_{rms-ps} = 18\mu$K)}
\end{figure}

Recent measurements of CBR anisotropy have provided some exciting results.
The large scale anisotropy detected by the COBE satellite allows us to
normalize the cosmological power spectrum at long wavelengths. The COBE
detection at a level of $\Delta T/T = 10^{-5}$ at 10$^\circ$ gives us
crucial information at scales above 10 degrees about the primordial
fluctuations. The largest scales do not however define the subsequent
evolution of the CBR structure in the collapse phase after decoupling. In
addition due to the limited number of sky patches available at large scales
along with the fact that we are only able to sample our local horizon and
not the entire universe (cosmic variance) limits the information available
from larger scale measurements. Additional measurements must be made at
smaller angular scales. As an example Figure 1 shows the power spectrum
expected in a number of models.

\section{CBR Anisotropy Measurements}

\begin{figure}
\vspace{11cm}  
\caption{Relevant backgrounds for terrestrial measurements
at the South Pole and at 35~Km where ACME observes. Representative galactic
backgrounds are shown for synchrotron, Bremsstrahlung and interstellar dust
emission as well as the various ACME (center) wavelength bands.}
\end{figure}

The spectrum of the cosmic background radiation peaks in the millimeter-wave
region. Figure 2 shows a plot of antenna temperature vs. frequency,
demonstrating the useful range of CBR observation frequencies and the
various backgrounds involved. The obvious regime for CBR measurements is in
the microwave and millimeter-wave regions.

In the microwave region, the primary extra-terrestrial foreground
contaminants are galactic synchrotron and thermal bremsstrahlung emission.
Below 50 GHz, both of these contaminants have significantly different
spectra than CBR fluctuations. Because of this, multi-frequency measurements
can distinguish between foreground and CBR fluctuations (provided there is
large enough signal to noise). For example, Figure 2 shows the ACME bands.

Above 50 GHz, the primary contaminant is interstellar dust emission. At
frequencies above 100 GHz, dust emission can be distinguished from CBR
fluctuations spectrally, also using multi-frequency instruments.

At all observation frequencies, extra-galactic radio sources are a concern.
For an experiment with a collecting area of 1m$^2$ (approximately a 0.5$%
^\circ$ beam at 30 GHz for sufficiently under-illuminated optics), a 10 mJy
source will have an antenna temperature of about 10 $\mu$K, which will produce
a significant signal in a measurement with a sensitivity of $\Delta T/T
\approx 1\times 10^{-6}$. Extra-galactic radio sources have the disadvantage
that there is no well known spectrum which describes the whole class. For
this reason, measurements over a very large range of frequencies and angular
scales are required for CBR anisotropy measurements in order to achieve a
sensitivity of $\Delta T/T \approx 1 \times 10^{-6}$.

\section{Instrumental Considerations}

Sub-orbital measurements differ from orbital experiments in at least one
important area, namely our terrestrial atmosphere is a potential
contaminant. A good ground-based site like the South Pole has an atmospheric
antenna temperature of 5~K at 40~GHz, for example. For a measurement to
reach an error of $\Delta T/T \approx 1 \times 10^{-6}$, the atmosphere must
remain stable over 6 orders of magnitude. In addition to this, the
atmosphere will contribute thermal shot noise. At balloon altitudes,
atmospheric emission is 3-4 orders of magnitude lower and much less of a
concern. In addition, the water vapor fraction is extremely low at balloon
altitude. Satellite measurements avoid this problem altogether. Another
consideration for CBR anisotropy measurements is the sidelobe antenna
response of the instrument. Astronomical and terrestrial sources away from
foresight can contribute significant signals if the antenna response is not
well behaved. Under-illuminated optical elements and off-axis low blockage
designs are typically employed for the task. The sidelobe pattern can be
predicted and well controlled with single-mode receivers, but appears to be
viable for multi-mode optics as well. Even with precautions, sidelobe
response will remain an area of concern for all experiments.

Most of the measurements to be discussed are limited by receiver noise when
atmospheric seeing was not a problem. It is possible to build receivers
today with sensitivities of 200-400~$\mu$K $\sqrt{s}$ using HEMTs or
bolometers. A balloon flight obtaining 10 hours of data on 10 patches of
sky, for example, could achieve a 1~$\sigma$ sensitivity of 6.7~$\mu$K or $%
\Delta T/T = 2.5\times 10^{-6}$ {\it per pixel} using one such detector.

To map CBR anisotropy with a sensitivity of $\Delta T/T=1\times 10^{-6}$
requires more integration time, lower noise receivers or multiple receivers.
A 14-day, long duration balloon flight launched from Antarctica could result
in a per pixel sensitivity of $\Delta T/T=5\times 10^{-7}$, if 10 patches
could be observed with a single detector element or $\Delta T/T=5\times
10^{-6}$ on 1000 patches as another example. Multiple detectors obviously
help here.

Measurements from the South Pole are also very promising. The large
atmospheric emission (compared to the desired sensitivity level - few
million times larger!) is of great concern and based upon our experience,
even in the best weather, there is significant atmospheric noise. Estimated
single difference atmospheric noise with a 1.5 degree beam is about 1~mK$
\sqrt{s}$ at 30 GHz during the best weather. This added noise, as well as
the overall systematic atmospheric fluctuations, make ground-based
observations challenging but so far possible, and, in fact, yielding the
most sensitive results.

Another approach to the problem is to use very low noise receivers and
obtain the necessary integration time by flying long duration balloons.
These receivers can be tested from ground-based observing sites like the
South Pole. Should the long duration balloon effort prove inadequate, the
only means toward the goal of mapping CBR anisotropy at this level will be a
dedicated satellite. Again, the receivers on such a satellite would have to
be low noise. The minimal cryogenic requirements for HEMT (High Electron
Mobility Transistor) amplifiers make them an obvious choice for satellite
receivers, but bolometric receivers using ADR coolers or dilution
refrigerators offer significant advantages at submillimeter wavelengths.

\section{History of the ACME Experiments}

In 1983, with the destruction of the 3~mm mapping experiment (Lubin et al.
1985), we decided to concentrate on the relatively unexplored degree scale
region. Motivated by the possibility of discovering anisotropy in the
horizon scale region where gravitation collapse would be possible and with
experience using very low noise coherent detectors at balloon altitudes, we
started the ACME program. A novel optical approach, pioneered at Bell
Laboratories for communications, was chosen to obtain the extreme sidelobe
rejection needed. In collaboration with Robert Wilson's group at Bell Labs,
a 1 meter off-axis primary was machined. A lightweight, fully-automated,
stabilized, balloon platform capable of directing the 1 meter off-axis
telescope was constructed. As the initial detector we chose a 3~mm SIS
receiver. Starting with lead alloy SIS junctions and GaAs FET pre-amplifiers
we progressed to Niobium junctions and a first generation of HEMTs to
achieve chopped sensitivities of about 3~mK~$\sqrt{s}$ in 1986 with
a beam size of 0.5 degrees FWHM at 3~mm.

The first flight was in August 1988 from Palestine, Texas. Immediately
afterwards, ACME was shipped to the South Pole for ground-based
observations. The results were the most sensitive measurements to date (at
that time) with 60~$\mu$K errors per point at 3~mm. The primary advantage of
the narrow band coherent approach is illustrated in Figure 2 where we plot
atmospheric emission versus frequency for sea level, South Pole (or 4~km
mountain top) and 35~km balloon altitudes. With a proper choice of
wavelength and bandpass, extremely low residual atmospheric emission is
possible. (Total $<$ 10~mK. The differential emission, over the beam throw,
is much smaller.) Another factor of 10 reduction is possible in the
``troughs" in going to 40~km altitude. The net effect is that atmospheric
emission does not appear to be a problem in achieving $\mu$K level
measurements, if done appropriately.

Subsequently, ACME has been outfitted with a variety of detector including
direct amplification detectors using HEMT technology. These remarkable
devices developed largely for communications purposes are superb at
cryogenic temperatures as millimeter wavelength detectors. Combining
relatively broad bandwidth (typically 10-40\%) with low noise
characteristics and moderate cooling requirements (including operation at
room temperature) they are a good complement to shorter wavelength
bolometers allowing for sensitive coverage from 10~GHz to 200~GHz when both
technologies are utilized. The excellent cryogenic performance is due in
large part to the efforts of the NRAO efforts in amplifier design
(Pospieszalski 1990). We have used both 8-12~mm and 6-8~mm HEMT detectors on
ACME, these observations being carried out from the South Pole in the 1990
and 1993 seasons. The beam sizes are 1.5 degrees and 1 degree FWHM for the
8-12 and 6-8~mm HEMTs respectively. Detectors using both GaAs and InP
technology have been used. The lowest noise we have achieved to date is 10~K
at 40~GHz, this being only 3.5 times the quantum limit at this frequency.
These devices offer truly remarkable possibilities. Figure 3 shows the basic
experiment configuration.

There have been a total of eleven ACME observations/flights from 1988
to 1994. Over twenty articles and proceedings have resulted from
these measurements as well as seven Ph.D. theses. A summary of the various
observations is given in Table I.

\begin{figure}
\vspace{7cm}  
\caption{Basic ACME configuration}
\end{figure}

\begin{table}[htb]
\begin{center}
\caption{CBR Measurements made with ACME}
\begin{tabular}{lllll}
\hline
&&& Beam & \\
 Date  &  Site &  Detector System & FWHM & \quad Sensitivity\\
&&& (deg) & \\
\hline
1988 Sep & Balloon$\scriptstyle{^{\rm P}}$  & 90 GHz SIS receiver & 0.5 & 4 mK
s$\scriptstyle{^{1/2}}$\\
1988 Nov-1989 Jan & South Pole & 90 GHz SIS receiver & 0.5 & 3.2\\
1989 Nov & Balloon$\scriptstyle{^{\rm FS}}$ & MAX photometer
(3, 6, 9, 12 cm$\scriptstyle{^{-1}}$) $\scriptstyle{^3}$He & 0.5 & 12, 2, 5.7,
7.1\\
1990 Jul & Balloon$\scriptstyle{^{\rm P}}$ & MAX photometer
(6, 9, 12 cm$\scriptstyle{^{-1}}$) $\scriptstyle{^3}$He & 0.5 & 0.7, 0.7, 5.4\\
1990 Nov-1990 Dec & South Pole & 90 GHz SIS receiver & 0.5 & 3.2\\
1990 Dec-1991 Jan & South Pole & 4 Channel HEMT amp (25-35 GHz) & 1.5 & 0.8\\
1991 Jun & Balloon$\scriptstyle{^{\rm P}}$ & MAX photometer
(6, 9, 12 cm$\scriptstyle{^{-1}}$) $\scriptstyle{^3}$He & 0.5 & 0.6, 0.6, 4.6\\
1993 Jun & Balloon$\scriptstyle{^{\rm P}}$ & MAX photometer (3, 6, 9, 12
cm$\scriptstyle{^{-1}}$) ADR &
0.55-0.75 & 0.6, 0.5, 0.8, 3.0\\
1993 Nov-1994 Jan & South Pole & HEMT 25-35 GHz & 1.5 & 0.8\\
1993 Nov-1994 Jan & South Pole & HEMT 38-45 GHz & 1.0 & 0.5\\
1994 Jun & Balloon$\scriptstyle{^{\rm P}}$ & MAX photometer (3, 6, 9, 14
cm$\scriptstyle{^{-1}}$) ADR &
0.55-0.75 & 0.4, 0.4, 0.8, 3.0\\
\hline
\end{tabular}
\end{center}
\end{table}
\vspace{-5mm} {\noindent\hsize=6.5in Sensitivity does not include atmosphere
which, for ground-based experiments, can be substantial. }

\vspace{5mm}

\noindent P\hspace{2mm} - Palestine, TX

\noindent FS - Fort Sumner, NM

\section{The MAX Experiment}

During the construction of ACME, a collaboration was formed between our
group and the Berkeley group (Richards/Lange) to fly bolometric detectors on
ACME. This fusion is called the MAX experiment and subsequently blossomed
into the extremely successful Center for Particle Astrophysics' CBR effort.
Utilizing the same basic experimental configuration as other ACME
experiments, MAX uses very sensitive bolometers from about 1-3~mm wavelength
in 3 or 4 bands. Flown from an altitude of 35~km, MAX has had five very
successful flights. The first MAX flight (second ACME flight) occurred in
June 1989 using $^3$He cooled (0.3~K) bolometers, and the most recent flight
occurred in June 1994 using ADR (Adiabatic Demagnetization Refrigeration)
cooled bolometers. All the MAX flights have had a beam size of near 0.5
degrees.

\section{Evidence for Structure Prior to COBE}

Prior to the COBE launch, ACME had made two flights and one South Pole
expedition. Prior to the April 92 COBE announcement, ACME had flown four
times and made two South Pole trips for a total of seven measurements. Our
1988 South Pole trip with ACME outfitted with a sensitive SIS
(Superconductor-Insulator-Superconductor) receiver resulted in an upper
limit of $\Delta T/T \ltwid 3.5\times 10^{-5}$ at $0.5^{\circ }
$ for a Gaussian sky. This was tantalizingly close to the ``minimal
predictions" of anisotropy at the time and as we were to subsequently
measure, just barely above the level of detectability. In the fall of 1989,
we had our first ACME-MAX flight with a subsequent flight the next summer
(so called MAX-II flight). Remarkably, when we analyzed the data from this
second flight, we found evidence for structure in the data consistent with a
cosmological spectrum. This was data taken in a low dust region and showed
no evidence for galactic contamination. The data in the Gamma Ursa Minoris
region (``GUM data") was first published in Alsop et al. (1992)
\underbar{\bf prior} to the announcement of the COBE detections. At the time
our most serious concern was of atmospheric stability so we decided to revisit
this region in the next ACME flight in June 1991. In the meantime, ACME was
shipped to the South Pole in October 1990 for another observing run, this time
with both an SIS detector and a new and extremely sensitive HEMT receiver.
At scales near 1 degree, close to the horizon size, results from the South
Pole using the ACME (Advanced Cosmic Microwave Experiment) with a High
Electron Mobility Transistor (HEMT) based detector placed an upper limit to
CBR fluctuations of $\Delta T/T\leq 1.4\times 10^{-5}$ at 1.2$^{\circ }$
(Gaier et al. 1992). This data set has significant structure in excess of
noise with a spectrum that was about 1.4$\sigma$ from flat
(Gaier 1993). This upper limit for a Gaussian autocorrelation function sky was
computed from the highest frequency channel. Since the data is taken in a step
scan and not as a continuous scan it is not possible to eliminate the
possibility that the structure seen is cosmological since the beam size varies
from channel to channel. Under the assumption that the structure seen is
cosmological, a four channel average of the bands yields a detection at the
level of $\Delta T/T\simeq1\times 10^{-5}$ (Bond 1993). Interestingly, this is
about the same level seen in another SP 91 scan (see next) as well as that
observed in the nearly same region of sky observed in the SP 94 data
(Gundersen et al. 1995).

Additional analysis of the 1991 South Pole data using another region of
the sky and with somewhat higher sensitivity shows a significant detection
at a level of $\Delta T/T=1\times 10^{-5}$ (Schuster et al. 1993). The
structure observed in the data has a relatively flat spectrum which is
consistent with CBR but could also be Bremsstrahlung or synchrotron in
origin. This data also sets an upper limit comparable to the Gaier et al.
upper limit, but can also be used to place a lower limit to CBR fluctuations
of $\Delta T/T\geq 8\times 10^{-6}$, if all of the structure is attributed
to the CBR. The 1$\sigma$ error measured per point in this scan is 14 $\mu$K
or $\Delta T/T=5\times 10^{-6}$. Per pixel, this is the most sensitive CBR
measurement to date at any angular scale. Combining these two scans in a
multichannel analysis results in a detection level slightly above $1\times
10^{-5}$ (Bond 1993).  The relevant measurements just prior to the COBE
announcement are summarized in Figure 4. With apparent detection and good
upper limits at degree scales, what was needed was large scale normalization.
This was provided by the COBE data in 1992 and, as shown in Figure 5, the
degree scale measurements were consistent with COBE given the errors
involved. Without the large scale normalization of the COBE data, it was
hard to reconcile the apparently discordant data. However, with the
refinement in theoretical understanding and additional data, the pre-COBE
ACME data now are seen to be remarkably consistent with the post-COBE data.

\begin{figure}
\vspace{11cm}  
\caption{ACME CBR Power Spectrum data prior to the COBE
detection. Theoretical curves are from Figure 1. See KEY in Figure 5 caption.}
\end{figure}

\begin{figure}
\vspace{11cm}  
\caption{Figure 5: Recent ACME Results (in BOLD) along with results from other
groups. Key: a-COBE, b-FIRS, c-Tenerife,  d1-SP91 9 pt. 4 channel
analysis-Bond 93, d3-SP91 9+13pt 4 channel analysis-Bond 93, d5-SP91 9 pt.
Gaier et al. 92, e-Big Plate, f-PYTHON, g-ARGO, h-MAX4-Iota Dra, i-MAX4-GUM,
j-MAX4-Sig Herc, k-MSAM2, l-MSAM2, m-MAX3-GUM, n-MAX3-mu Peg, o-MSAM3, p-MSAM3,
q-Wh. Dish, r-OVRO7, s2-SP94-Q, s3-SP94-Ka, t-SP89, u-MAX2-GUM, many from
Steinhardt and Bond, private communication.}
\end{figure}

\section{Results}

There have been a total of eleven ACME observations/flights from 1988 to
1994. ACME articles by Meinhold \& Lubin (1991), Meinhold et al. (1992),
ACME-HEMT articles by Gaier et al. (1992), Schuster et al. (1993), Gundersen
et al. (1995) and ACME-MAX articles by Fischer et al. (1992, 1995), Alsop
et al. (1992), Meinhold et al. (1993), Gundersen et al. (1993), Devlin et al.
1994 and Clapp et al. 1994 summarize the results to date.

Significant detection by ACME at 1.5 degrees is reported by Schuster et al.
(1993) at the $1\times 10^{-5}$ level and by Gundersen et al. (1993) at 0.5
degrees at the $4\times 10^{-5}$ level in adjacent issues of {\sl ApJ Letters%
}. The lowest error bar per point of any data set to date is in the Schuster
et al. 1.5$^{\circ }$ data with 14~$\mu $K while the largest signal to noise
signal is in Gundersen et al.(1993) with about a 6~$\sigma $ detection (at the
peak). Recently Wollack et al. (1993) reported a detection at an angular
scale of 1.2 degrees of about $1.4\times 10^{-5}$ consistent with Schuster
et al., Gaier et al. and the combined 9 +\ 13 pt. analysis using a detector
and beam size  nearly identical to ours. Remarkably this result is taken in
a completely different region of the sky and at lower galactic latitude and
yields similar results. A conspiracy to yield comparable results in very
different parts of the sky from point sources or sidelobe spill over is
always possible. When one takes into account the similar level of the COBE
detection at larger scales it would seem to require multiple conspiracies
however. Additional data taken at the South Pole by ACME in 1993/94 (``SP94"
data) in a region of the sky close to the SP91 area but with additional HEMT
detectors from 25 - 45 GHz in seven bands and with beam sizes from 1.0 - 1.7
degrees FWHM, yield results
consistent with a CDM model (and others) normalized to COBE. The SP94 data
are consistent with the  SP91 results (Gundersen et al. 1995) as shown in
Figure 5. At 0.5 degrees, the MSAM group reports detection of a ``CBR
component" at a level of about $2\times 10^{-5}$ but with ``point like
sources" that are being reanalyzed and which may contribute additional power.
Our results from the June 1993 ACME-MAX flight give significant detections at
the $3-4\times 10^{-5}$ level at angular scales near 0.5 degrees.

The most recent ACME-MAX data have been in low dust regions so that no
subtraction of dust was needed. In one scan, the $\mu$-Pegasi region, there
was enough dust to provide a good calibration of  high galactic latitude
interstellar dust emission (Meinhold et al. 1993).  Interestingly, the residual
``CBR component" was anomolously low compared to  the other regions surveyed.
Whether this is indicative of other issues, such  as non-gaussian fluctuations,
or is just due to limited sampling statistics is unclear at this time.

The most recent ACME-MAX flight in June 1994 included two more low dust
regions and a revisit of the $\mu$-Pegasi region. The data is currently being
analyzed (Lim et al. 1995, Tanaka et al. 1995).

It is remarkable that over a broad range
of wavelengths, very different experiments using a variety of technologies and
observing in different parts of the sky report degree scale detection at the
one to a few $\times 10^{-5}$ level. ACME, in particular, has now been used to
measure
structure from 25 - 250 GHz and from 0.4 - 2 degrees that is in reasonably
good agreement with a CDM  power spectrum model. The agreement of the ACME-HEMT
data with other
experiments (notably Big Plate at Sasskatoon) up to $\ell$ of about 75 is very
good,
as can be seen from Figure 5. At 0.5 degree scales ($\ell$ about 150) the
agreement of ACME-MAX and MSAM data is marginal and will hopefully be
clarified soon. It is important to keep in mind that the statistical and
sampling errors need to be taken into account in any comparison between data
sets and between data and theory.

\begin{table}[htb]
\begin{center}
\caption{Recent ACME Degree Scale Results}
\begin{tabular}{lllcrc}
\hline
&& Beam & \\
Publication & Configuration & FWHM & $\Delta T/T\times 10^{-6}$
& $\ell$ & $C_\ell\ell(\ell+1)/2\pi^*$ \\
&& (deg) & (GACF)$^{**}$ && $\quad (\times 10^{-10})$\\
\hline
Meinhold \& Lubin 91 & ACME-SIS SP89 & 0.5 & $<35$ & 145 &
$<8.6$\\

Alsop et al. 92 & ACME-MAX-II (GUM) & 0.5 & $45^{+57}_{-26}$
& 143 & $9.6^{+13.7}_{-4.2}$\\

Gaier et al. 92 & ACME-HEMT SP91 & 1.5 & $<14$ & 58
& $<1.5$\\

Meinhold et al. 93 & ACME-MAX-III  & 0.5 & $<25$ & 143
&$<2.96$\\
& \quad ($\mu$ Peg - upper limit) \\

Meinhold et al. 93 & ACME-MAX-III & 0.5 & $15^{+11}_{-7}$
& 143 &\\
& \quad ($\mu$ Peg - detection) \\

Schuster et al. 93 & ACME-HEMT SP91 & 1.5 & $9^{+7}_{-4}$ & 58
&$0.76^{+0.80}_{-0.21}$\\

Bond 93 & SP91 4 channel 9+13 pt. Analysis & 1.5 && 58 &
$1.06^{+0.83}_{-0.29}$\\

Bond 93 & SP91 4 channel 9 pt. Analysis & 1.5 && 58 &
$0.5^{+0.80}_{-0.16}$\\

Gundersen et al. 93 & ACME-MAX-III (GUM) & 0.5 &
$42^{+17}_{-11}$ & 143 & $8.5^{+3.0}_{-2.2}$\\

Devlin et al. 94 & ACME-MAX-IV (GUM) & 0.55-0.75 &
$37^{+19}_{-11}$ & 129 & $6.1^{+3.9}_{-1.5}$\\

Clapp et al. 94 & ACME-MAX-IV & 0.55-0.75 & $33^{+11}_{-11}$
& 129 & $4.9^{+1.9}_{-1.4}$\\
& \quad (Iota Draconis) \\

Clapp et al. 94 & ACME-MAX-IV & 0.55-0.75 & $31^{+17}_{-13}$
& 129 & $4.3^{+3.0}_{-1.4}$\\
& \quad (Sigma Hercules) \\

Gundersen et al. 95 & ACME-HEMT SP94 & 1 & &
73 & $2.14^{+2.00}_{-0.66}$\\

Gundersen et al 95 & ACME-HEMT SP94 & 1.5 &  & 58 &
$1.17^{+1.33}_{-0.42}$\\

Lim et al. 94 & ACME-MAX-V & 0.5 & in progress \\

Tanaka et al. 94 & ACME-MAX-V  & 0.5 & in progress \\
\hline
\end{tabular}
\end{center}

$~~^*$ from P. Steinhardt \& R. Bond, priv. communication. 1$\sigma$ errors,
upper limits are 95\%

$^{**}$ GACF=Gaussian Autocorrelation Function - Upper limits and error bands
are 95\%

\end{table}

In any case, 1992 and 1993 were clearly historical years in cosmology and CBR
studies in particular. The ACME results along with the results of other
groups are shown in Figure 5. As can be seen by comparison to Figure 4 which
was the data prior to COBE, the pre- and post-COBE data are reasonably
consistent given the errors. The deluge of theoretical results and scrutiny
that followed COBE was a boon for degree scale results giving us a
theoretical insight we lacked just a few years ago. The current ACME degree
scale results are summarized in Table II.

\section{Detector Limitations - Present and Fundamental}

Detectors can be broadly characterized as either coherent or incoherent
being those that preserve phase or not, respectively. Masers, SIS and HEMTs
are coherent. Bolometers are incoherent. SIS junctions can also be run in an
incoherent video detector mode. Phase preserving detectors inherently must
obey an uncertainty relationship that translate into a minimum detector
noise that depends on the observation frequency, the so called quantum
limit. Incoherent detectors do not have this relationship but are ultimately
limited by the CBR background itself. At about 40~GHz, these fundamental
limits are comparable. Current detectors are not at these fundamental
limits, though they are within an order of magnitude for both HEMTs and
bolometers when used over moderate bandwidths. Here we include all effects
including coupling efficiencies. Currently both InP HEMTs and ADR and $^3$He
cooled bolometers exhibit sensitivities of under 500~$\mu$K s$^{1/2}$.
This assumes no additional atmospheric noise, true at balloon altitudes. For
ground-based experiments, even at the South Pole, atmospheric noise is
significant however.

Significant advances have been made in recent years in detector technology
with effective noise dropping by over an order of magnitude over the past
decade. With moderate bandwidths the fundamental limits for detectors are
about a factor of 5 below the current values, so fundamental technology
development is to be highly encouraged for both coherent and incoherent
detectors.

With current detectors, achieving 1~$\mu$K sensitivity requires roughly one
day per pixel for a single detector. This is appropriate for detector
limited, not atmospheric limited, detection. This would be appropriate for
balloon altitudes.

Small arrays of detectors are currently planned for several experiments.
This should allow $\mu$K per pixel sensitivity over 100 pixels in time
scales of a few weeks, suitable for long duration ballooning or polar
observations. If the fundamental detector limits could be achieved, the
effective time would drop to about a day. Factors of 2-3 reduction in
current detector noise are not unreasonable to imagine over the next five
years, and if they could be achieved, the above time scale would drop to
less than a week. Multiple telescopes are also possible. If we are willing
to accept a goal of 3~$\mu$K per pixel (1 part per million of the CBR)
instead of 1~$\mu$K then roughly 10 times as many pixels can be observed for
the same integration time allowing significant maps to be made from
balloon-borne detectors. A 10~$\mu$K error per pixel measurement would allow
100 times as many pixels to be measured in the same time. As we learn more
about the structure of the CBR and about the nature of low level foreground
emission the choice of sensitivity for a given angular scale will become
clearer.

\section{Spectrum Measurements}

A related area of interest that could yield interesting cosmology in the
next few years is the long wavelength spectrum. Although the spectrum of the
CBR has been extremely well characterized by the COBE FIRAS experiment in
the millimeter wavelength range. However, in the range of about 1-100~GHz,
where interesting physical phenomenon may distort the spectrum, much work
remains to be done; particularly, at the longest wavelengths. Fortunately,
the atmospheric emission is quite low over much of this range from both good
ground-based sites and extremely low at balloon altitudes. Galactic emission
and sidelobe contamination are of primary concern at the longest
wavelengths, but it is expected that a number of ground-based and possibly
balloon-borne experiments will be performed and should be encouraged.

A recent balloon-borne experiment, Schuster et al. (1994), is an example of
what might be done in the future from balloon spectrum experiments. With all
cryogenic optics and no windows, this experiment measured $T=2.71\pm 0.02$~K
at 90~GHz with negligible atmospheric contamination ($\sim$ a few mK) and no
systematic corrections. Errors of order 1~mK should be obtainable. The basic
configuration could be extended to longer wavelengths where much remains to
be done. In particular coherent measurements at 10 - 50 GHz from a balloon
could be done. The BLAST (Balloon Absolute Spectrometer)-ARCADE (Absolute
Radiometer for Cosmology, Astrophysics, and Diffuse Emission) experiment, a
joint UCSB-Goddard balloon borne experiment will attempt to exploit the low
atmospheric emission available from balloon altitudes using coherent HEMT
detectors in the 10-30 GHz range. Accuracies of under 1 milliKelvin may be
feasible. This will allow extremely sensitive measurements of long
wavelength distortions in the CBR should they exist. Since the spectral
deviation rises rapidly at long wavelengths as does the galactic emission
from synchrotron radiation, measurements in the 5-20 GHz range will be
particularly useful.

\section{Polarization}

Very little effort has been directed towards the measurement of the
polarization of the CBR compared to the effort in direct anisotropy
detection. In part, this is due to the low level of linear polarization
expected. Typically, the polarization is only 1-30\% of the anisotropy and
depends strongly on the model parameters (Steinhardt 1994). This is an area
which in theory can give information about the reionization history, scalar
and tensor gravity wave modes and large scale geometry effects. In the future,
this may be a very fruitful area of inquiry particularly at degree angular
scales.

\section{To Space}

The question of whether or not a satellite is needed to get the degree scale
``answer" is complex. There is no question that the measurements can be done
from space and given sufficient funding this is definitely the preferable
way. It is unclear at this time what the limitations from sub-orbital
systems will be and vigorous work is planned for sub-orbital platforms over
the next decade. The galactic and extragalactic background problem remains
the same for orbital and sub-orbital experiments. The atmosphere can be
dealt with, particularly from balloon-borne experiments, with careful
attention to band passes. Per pixel sensitivities in the $\mu$K region are
achievable with current and new technologies, HEMTs, and bolometers over
hundreds to thousands of pixels. The major issue will be control of
sidelobes and getting a uniform dataset. Ideally full sky coverage would be
best and this is one area where a long term space based measurement would be
ideal. In the control of sidelobe response a multi AU orbital satellite
would be a major advance. This advantage is lost for near Earth orbit
missions, however. European efforts such as SAMBA and COBRAS and US efforts
such as PSI, MAP and FIRE are examples of possible future space based
efforts. A low cost precursor mission such as the university led COFI
satellite is an example of an economical approach to proving HEMT technology
in space for a possible future effort. By the end of the millennium, degree
scale maps over a reasonable fraction of the sky at the $10^{-6}$ level
should be possible from balloons and the ground. The potential knowledge to
be gained is substantial, and I can think of few areas of science where the
potential ``payoff" to input (financial and otherwise) is so high.

\section{Acknowledgements}

This work was supported by the National Science Foundation Center for Particle
Astrophysics, the National Aeronautics and Space Administration, the NASA
Graduate Student Research Program, the National Science Foundation Division of
Polar Programs, the California Space Institute, the University of California,
and the U.S. Army. Its success is the result of the work of a number of
individuals, particularly the graduate students and post doc's involved, in
particular Peter Meinhold, Alfredo Chincquanco, Jeffery Schuster, Michael
Seiffert, Todd Gaier, Tim Koch, Joshua Gundersen, Mark Lim, John Staren, Thyrso
Villela, Alex Wuensche, and Newton Figueiredo. The bolometric portions of the
ACME program (MAX) were in collaborations with Paul Richards' and Andrew
Lange's
groups at UCB and in particular with Mark Fischer, David Alsop, Mark Devlin,
Andre Clapp and Stacy Tanaka. The entire ACME effort would not have been
possible without the initial support and vision of Nancy Boggess and Don
Morris. Dick Bond and Paul Steinhardt supplied much appreciated theoretical
input to the data analysis and interpretation. The exceptional HEMT amplifier
was provided by NRAO and in particular
by Marion Pospieszaski and Michael Balister. Robert Wilson, Anthony Stark, and
Corrado Dragone, all of AT\&T Bell Laboratories, provided critical support and
discussion regarding the early design of the telescope and receiver system as
well as providing the primary mirror. I would like to thank Bill Coughran and
all of the South Pole support staff for highly successful 88-89, 90-91 and
93-94 polar summers. In addition, I want to acknowledge the crucial
contributions of the entire team of the National Scientific Balloon Facility in
Palestine, Texas for their continued excellent support. Finally, I would also
like to
thank my wife Georganne for providing the loving support to make this program a
reality.

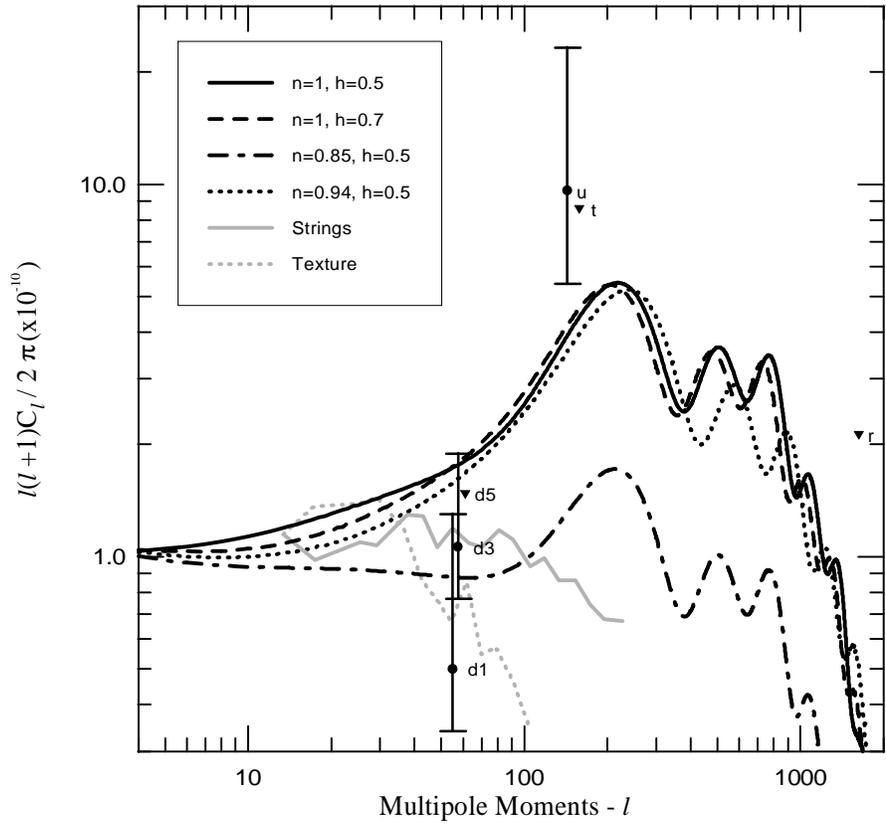
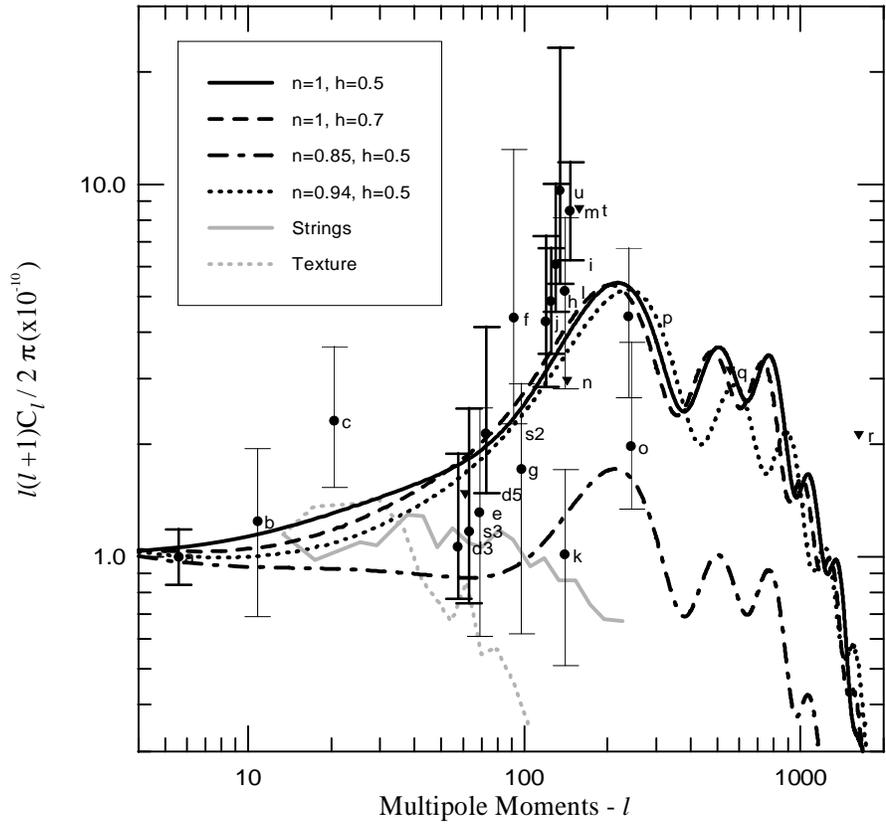
\end{document}